\documentclass[reprint,superscriptaddress,amsmath,amssymb,aps,prb,showpacs]{revtex4-2}
\usepackage{mhchem}
\usepackage{physics,slashed}
\usepackage{amsfonts,amsmath,amssymb}
\usepackage{mathrsfs}
\usepackage{bm}

\usepackage{graphicx}   
\usepackage{verbatim}   
\usepackage{color}      
\usepackage{hyperref}  
\usepackage{url}
\definecolor{nicered}{rgb}{0.5,0.,0.}
\definecolor{nicegreen}{rgb}{0.,0.5,0.}
\definecolor{niceblue}{rgb}{0.,0.,0.5}
\hypersetup{colorlinks,citecolor=nicegreen,linkcolor=nicered,urlcolor=niceblue}
\usepackage{array}
\usepackage{dcolumn}
\usepackage{multirow}
\usepackage{cprotect}
\usepackage{framed}
\usepackage{ragged2e}

\begin{document}
\title{Stoichiometry-Controlled Structural Order and Tunable Antiferromagnetism in $\mathrm{Fe}_{x}\mathrm{NbSe_2}$ ($0.05 \le x \le 0.38$)}

\author{Xiaotong Xu}
\affiliation{Tsung-Dao Lee Institute, School of Physics and Astronomy, Shanghai Jiao Tong University, Shanghai 200240, China}

\author{Bei Jiang}
\affiliation{Tsung-Dao Lee Institute, School of Physics and Astronomy, Shanghai Jiao Tong University, Shanghai 200240, China}

\author{Runze Wang}
\affiliation{Tsung-Dao Lee Institute, School of Physics and Astronomy, Shanghai Jiao Tong University, Shanghai 200240, China}

\author{Zhibin Qiu}
\affiliation{Shenzhen Institute for Quantum Science and Engineering, Southern University of Science and Technology, Shenzhen, 518055, China}
\affiliation{International Quantum Academy, Shenzhen 518048, China.}

\author{Shu Guo}
\affiliation{Shenzhen Institute for Quantum Science and Engineering, Southern University of Science and Technology, Shenzhen, 518055, China}
\affiliation{International Quantum Academy, Shenzhen 518048, China.}

\author{Baiqing Lv}
\affiliation{Tsung-Dao Lee Institute, School of Physics and Astronomy, Shanghai Jiao Tong University, Shanghai 200240, China}
\affiliation{Zhangjiang Institute for Advanced Study, Shanghai Jiao Tong University, Shanghai 200240, China}
\affiliation{State Key Laboratory of Micro-nano Engineering Science, Shanghai Jiao Tong University, Shanghai 200240, China}

\author{Ruidan Zhong}
\thanks{E-mail:rzhong@sjtu.edu.cn}
\affiliation{Tsung-Dao Lee Institute, School of Physics and Astronomy, Shanghai Jiao Tong University, Shanghai 200240, China}	
\affiliation{State Key Laboratory of Micro-nano Engineering Science, Shanghai Jiao Tong University, Shanghai 200240, China}

\date{\today}
	
\begin{abstract}
Transition metal dichalcogenides (TMDs) enable magnetic property engineering via intercalation, but stoichiometry-structure-magnetism correlations remain poorly defined for Fe-intercalated $\mathrm{NbSe_2}$. Here, we report a systematic study of $\mathrm{Fe}_{x}\mathrm{NbSe_2}$ across an extended composition range $0.05 \le x \le 0.38$, synthesized via chemical vapor transport and verified by rigorous energy-dispersive X-ray spectroscopy (EDS) microanalysis. x-ray diffraction, magnetic, and transport measurements reveal an intrinsic correlation between Fe content, structural ordering, and magnetic ground states. With increasing $x$, the system undergoes a successive transition from paramagnetism to a spin-glass state, then to long-range antiferromagnetism (AFM), and ultimately to a reentrant spin-glass phase, with the transition temperatures exhibiting a nonmonotonic dependence on Fe content. The maximum Néel temperature ($T_{\mathrm{N}}$ = $\mathrm{175K}$) and strongest AFM coupling occur at $x=0.25$, where Fe atoms form a well-ordered $2a_0 \times 2a_0 $ superlattice within van der Waals gaps. Beyond $x = 0.25$, the superlattice transforms or disorders, weakening Ruderman-Kittel-Kasuya-Yosida (RKKY) interactions and significantly reducing $T_{\mathrm{N}}$. Electrical transport exhibits distinct anomalies at magnetic transition temperatures, corroborating the magnetic state evolution. Our work extends the compositional boundary of Fe-intercalated $\mathrm{NbSe_2}$, establishes precise stoichiometry-structure-magnetism correlations, and identifies structural ordering as a key tuning parameter for AFM. These findings provide a quantitative framework for engineering altermagnetic or switchable antiferromagnetic states in van der Waals materials.	
	\end{abstract}
	
\maketitle

\section{Introduction}~\label{sec:A}
Transition metal dichalcogenides (TMDs) represent a versatile class of layered materials characterized by weak van der Waals interactions between covalently bonded atomic layers~\cite{CSR2021TMDreview, MT2017TMDreview},
which enables unique opportunities for intercalation chemistry~\cite{SM2020ITMDreview, IRPC1983ITMDreview}.
Guest species engineer properties of the host by affecting the interlayer interactions and inducing charge transfer~\cite{NRC2024ITMDreview}.
Among various intercalated TMDs, $3d$-transition metal intercalated TMDs have attracted significant attention due to their composition-dependent magnetic properties and phase transitions~\cite{PRB2006FeTiSe2, CM2002AMoS2, PRL2023FexNbS2, PRB2024CoTaSe2, PRB2020Mn0.25NbS2}.
Magnetic ground states of systems undergo dramatic evolution with concentration~\cite{JAP1981FexTaS2, PRB2016FexTaS2, JSCC1994FexTaS2}.
This extreme sensitivity to precise stoichiometry highlights the importance of accurate composition control and measurement when investigating physical properties in intercalated TMDs.

\ce{2H-NbSe2} is an excellent host for intercalation due to its distinctive electronic properties.
It crystallizes in a hexagonal lattice with niobium atoms sandwiched between selenium layers in trigonal prismatic coordination, creating gaps for accommodating foreign species.
\ce{2H-NbSe2} exhibits both superconductivity below 7.2 K and a charge density wave transition at approximately 33 K, providing an ideal platform for studying the relationship between structural ordering and emergent physical phenomena in two-dimensional materials through intercalation~\cite{PRB1972NbSe2, PRL2011NbSe2, NL2018NbSe2}.
$3d$-transition metal intercalated $\mathrm{NbSe_2}$ compounds show significant modifications to their electronic and magnetic properties~\cite{PRB2021CrNbSe2, SS2025MnNbSe2, PRB1973ANbSe2, PRB2025NiNbSe2}.
Notably, $\mathrm{Co_{1/4}NbSe_2}$ has been reported to display A-type antiferromagnetism (AFM) and spin-split bands~\cite{NC2025CoNbSe2}.
Interlayer Co atoms with specific stoichiometries $x = 1/4$ form an ordered $2 a_0 \times2 a_0 $ superstructure with special periodicity, resulting in time reversal symmetry breaking (TRSB) in the material.
This TRSB characteristic defines an AFM state known as altermagnetism~\cite{JPSJ2019AM, PRX2022AM, PRB2023AMMT}, which exhibits considerable implications for future spin-based electronic applications.
Fe-intercalated analogous system $\mathrm{Fe}_{x}\mathrm{NbSe_2}$, which also displays AFM ordering, represents a promising candidate for altermagnetic materials, particularly at the stoichiometric ratio of $x = 1/4$. By analogy, \ce{Fe_1/4NbSe2} is a candidate altermagnet, yet its stoichiometric stability and magnetic robustness remain unexplored across extended doping ranges.

Previous investigations of Fe-intercalated $\mathrm{NbSe_2}$ have been predominantly focused on the low-concentration regime $x\mathrm{\le1/4}$~\cite{JPCC2023Bridging, CPB2025Fe0.25NbSe2}.
However, the extreme sensitivity of physical properties to minor variations in Fe concentration necessitates precise stoichiometric characterization.
Many studies have employed only nominal ratios of initial reactants as Fe stoichiometry without compositional verification, risking misinterpretation of physical phenomena due to undetected stoichiometric discrepancies between intended and actual compositions~\cite{PRB1979FexNbSe2, PRB1993STM, JJAP2013FeNbSe2}.

In this paper, we present a systematic investigation of $\mathrm{Fe}_{x}\mathrm{NbSe_2}$ across a broad range of iron concentrations $0.05 \le x \le 0.38$.
By establishing accurate correlations between verified Fe content and observed physical properties, our studies reveal the evolution of magnetic ordering from paramagnetic (PM) behavior to a spin-glass state, then to AFM and eventually back to a spin-glass state, with transition temperatures exhibiting notable dependence on Fe content that follows distinct trends near commensurate concentrations.
Electrical properties of materials have also been modified, evidencing the transformation of magnetic ordering.

\begin{figure*}[tbp]
\includegraphics[width = 0.9\textwidth]{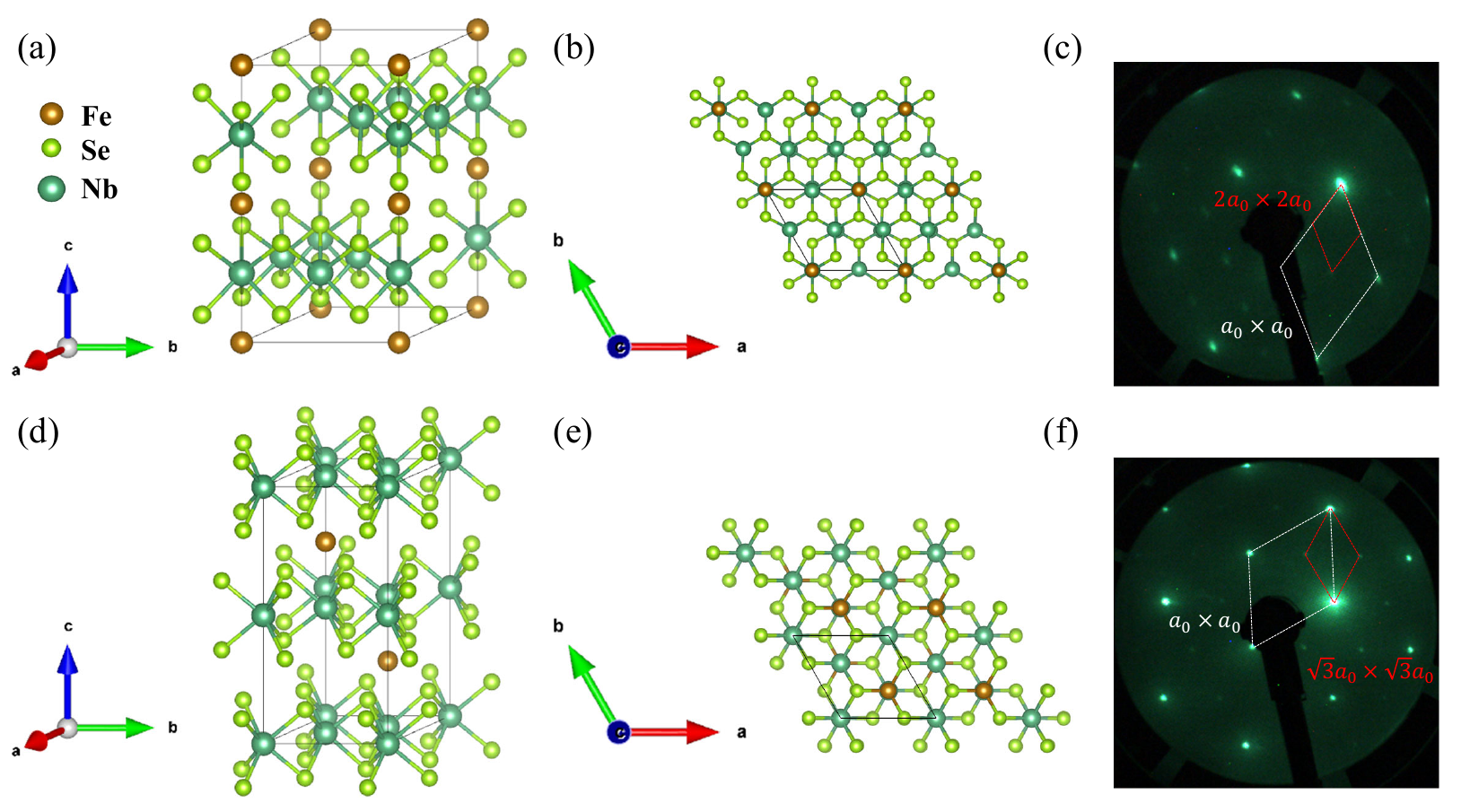}
\caption{\justifying
Structural and surface signatures of commensurate superstructures in \ce{Fe_{x}NbSe2}. (a),(d) Side views of the crystal structure for $x = 1/4$ and $x = 1/3$, respectively. (b),(e) Top-down projections along the c axis showing Fe atoms (orange) at ordered positions in \ce{Fe_{1/4}NbSe2} and \ce{Fe_{1/3}NbSe2}, respectively. (c),(f) Corresponding LEED patterns recorded at 105 eV and 115 eV, respectively. The red overlays denote the enlarged reciprocal-unit cells, directly confirming the real-space superstructures induced by Fe ordering in the van der Waals gaps.
}
\label{fig:str}
\end{figure*}

\section{Experiment}~\label{sec:B}

Single crystals of $\mathrm{Fe}_{x}\mathrm{NbSe_2}$ were synthesized via the chemical vapor transport method~\cite{JPCC2022Evolution,JPCC2023Bridging}.
Stoichiometric ratios of Nb powder (99.99\%), Se shots (99.9999\%), and Fe powder (99.5\%) were weighed, ground in an agate mortar and pestled under the argon atmosphere, along with the $\mathrm{3~ mg~ cm^{-1}}$ iodine.
Mixtures were vacuum-sealed in a quartz tube and placed in a horizontal two-zone furnace.
Typically, a thermal gradient of $\mathrm{800 ~^{\circ} C} $ to $\mathrm{700 ~^{\circ} C} $ was established at a ramping rate of $ \mathrm{ 3~^{\circ}C ~min^{-1}} $ and maintained for 14 days. Then the quartz tube was cooled down to room temperature at a rate of $ \mathrm{ 3~^{\circ}C ~min^{-1}} $.
All major compositions reported in this study followed this protocol, unless otherwise noted. Minor variations in the synthesis conditions for specific samples are discussed in detail in the Results and Discussion section.
Upon cooling operation, millimeter-sized crystalline platelets were observed in the tube. The crystals underwent repeated washing with ethanol to eliminate residual surface iodine.

To confirm the actual proportion of Fe in the samples, energy-dispersive x-ray spectroscopy (EDS) analyses were conducted on a Sigma 300 instrument with an EDS detector.
At least three specimens from each batch were tested, with at least three sites randomly selected and analyzed per specimen.
The actual ratios of Fe in the corresponding samples were determined by averaging the results obtained from multiple mapping-scanning EDS measurements.

Structural characterization of samples with varying intercalation ratios ($x$) was performed using x-ray diffraction (XRD).
Powder XRD data were collected at room temperature from $10^{\circ}$ to $90^{\circ}$ using a Bruker D8 Advance Eco diffractometer equipped with $\mathrm{Cu-K\alpha}$ radiation ($\mathrm{\lambda = 1.5418 \mathring{A}}$).
Rietveld refinement~\cite{Rietveld:a07067} of the powder diffraction patterns was conducted using TOPAS v.6.0 software. To examine whether samples with specific compositions form a superstructure, low energy electron diffraction (LEED) measurements were performed using an LEED 800 system (OCI Vacuum Microengineering). The samples were cleaved \textit{in situ} at 70 K under ultrahigh vacuum with a base pressure better than $1 \times 10^{-10}$ Torr.

Temperature-dependent dc magnetic susceptibility was determined on single-crystal samples under a magnetic field via a Physical Property Measurement System (PPMS, Quantum Design) equipped with a vibrating sample magnetometer option. A 0.1 T magnetic field was applied for spin-glass sample measurements, while 1 T was used for antiferromagnetic samples.
A magnetic field is applied parallel to the c axis of the samples.
Temperature dependence of the resistivity was measured via PPMS at zero field, employing the standard four-probe method.
The measurements were performed over a temperature region from 2 K to 300 K. Electrical contacts were fabricated using Au wires and silver paint.

\section{Results and Discussion}~\label{sec:C}
\subsection{Structural characterization of crystals}\label{sec:C1}

\begin{figure*}[tbp]
\centering
\includegraphics[width = 0.8\textwidth]{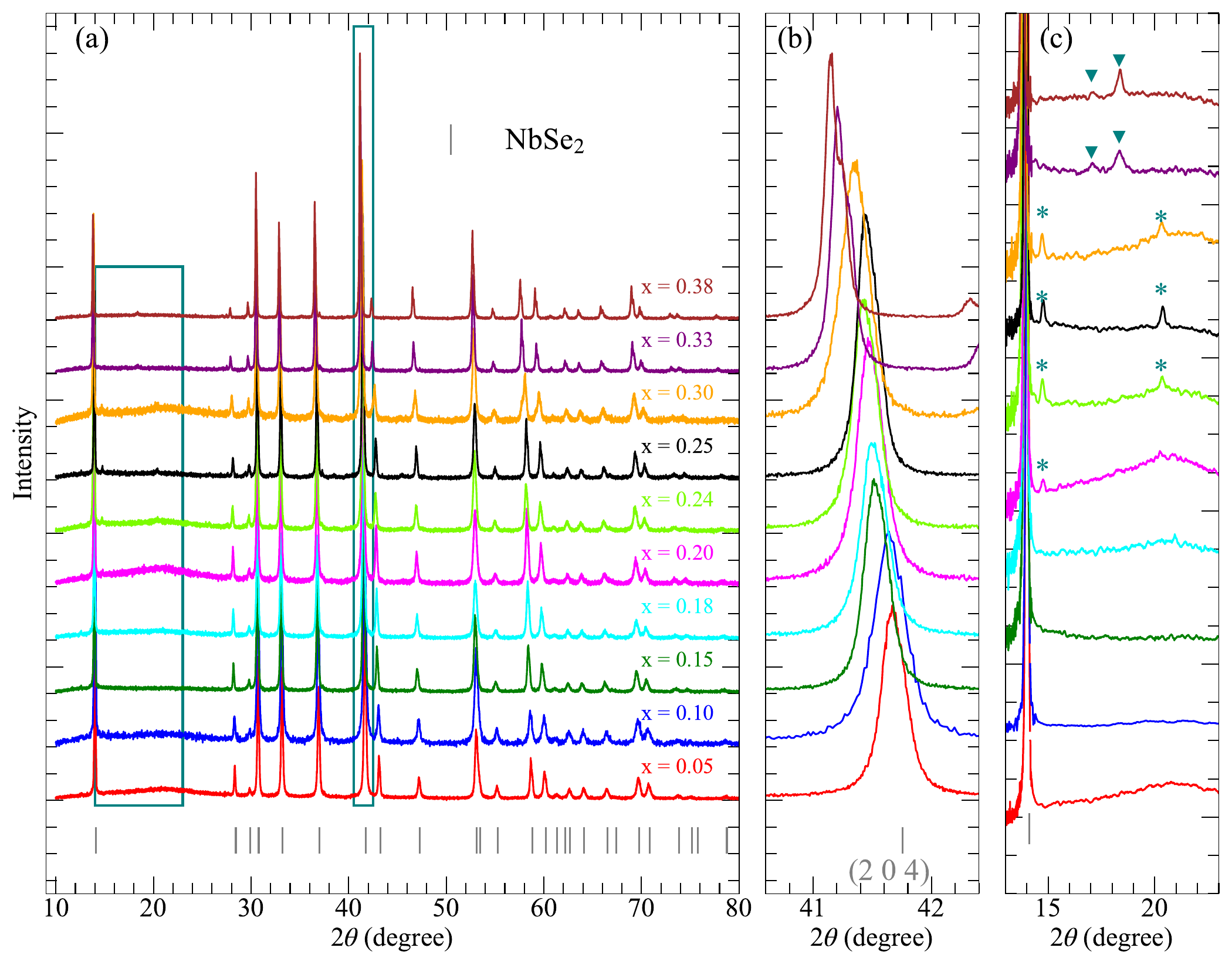}
\caption{
(a) Powder x-ray diffraction patterns of \ce{Fe_{x}NbSe2} ($0.05 \leq x \leq 0.38$) at room temperature. All samples crystallize in the hexagonal \ce{2H-NbSe2} structure. (b) Magnified view of the (204) peak showing a systematic shift to lower $2\theta$ with increasing Fe content $x$. (c) Low-angle region displaying superlattice peaks arising from the $2a_0 \times 2a_0$ (stars) and $\sqrt{3}a_0 \times \sqrt{3}a_0$ (inverted triangles) superstructure.
}
\label{fig:pxrd}
\end{figure*}

For synthesis, the initial ratio of reactants and test results of samples are presented in Table~\ref{tab:eds}. The Fe concentrations investigated in this work covered the range of $0.05 \le x \le 0.38$.

For samples with $x > 0.15$, the actual iron content was consistently lower than the initial Fe ratio in the starting mixture, suggesting excess iron likely formed iron-iodide compounds during the slow cooling process of crystal growth or remained as unreacted reactants at the high-temperature zone.
The inhomogeneity among different specimens from the same batch can be negligible, as they share consistent EDS results.
Due to the affinity of Fe for iodine, even with identical initial reactant ratios (e.g., $x$=0.33), increasing the iodine amount from $\mathrm{3~ mg~ cm^{-1}}$ to $\mathrm{5~ mg~ cm^{-1}}$ leads to a lower Fe content as in $\mathrm{Fe}_{0.20}\mathrm{NbSe_2}$.
Raising the temperature of both zones by~$50\,^{\circ}\mathrm{C}$ suppresses Fe intercalation, such that even with a high nominal Fe ratio of~0.75 in the reactants, the resulting crystals exhibit low actual Fe content.
The experimental results indicate that the maximum Fe content in the single-crystal samples was 0.38, corresponding to an initial molar ratio of Fe:Nb:Se set at 0.60:1:2.
Subsequent optimization experiments failed to prepare samples with higher Fe content, demonstrating the stoichiometric limit for Fe intercalation under these synthesis conditions.

\begin{table}[htbp] 
  \centering
  \caption{The initial ratio of reactants and elemental composition, with corresponding $x$ values of samples measured with EDS. Fe content was calculated relative to Nb, and results are obtained by calculating the mean of multiple measurements.}
  \label{tab:eds}
  \begin{tabular}{cc}
    \toprule
    Initial (Fe:Nb:Se) & EDS (Nb as 1) \\
\hline
    $0.05:1:2$ & $\mathrm{Fe_{0.05}NbSe_{2.03}}$ \\
    $0.10:1:2$ & $\mathrm{Fe_{0.10}NbSe_{2.05}}$ \\
    $0.15:1:2$ & $\mathrm{Fe_{0.15}NbSe_{2.04}}$ \\
    $0.20:1:2$ & $\mathrm{Fe_{0.18}NbSe_{2.07}}$ \\
    $0.33:1:2$ & $\mathrm{Fe_{0.20}NbSe_{1.96}}$ \\
    $0.33:1:2$ & $\mathrm{Fe_{0.24}NbSe_{1.76}}$ \\
    $0.40:1:2$ & $\mathrm{Fe_{0.25}NbSe_{1.92}}$ \\
    $0.75:1:2$ & $\mathrm{Fe_{0.30}NbSe_{1.92}}$ \\
    $0.50:1:2$ & $\mathrm{Fe_{0.33}NbSe_{1.77}}$ \\
    $0.60:1:2$ & $\mathrm{Fe_{0.38}NbSe_{1.83}}$ \\
\hline
  \end{tabular}
\end{table}

Fe atoms intercalated into $\mathrm{NbSe_2}$ occupy octahedral interstitial sites within the van der Waals gap, same as intercalation of other $3d$-transition metals in TMDs~\cite{JSSC1970, NL2023MVSe2, PRB2025CuxTaS2}.
At lower concentrations $(x \le 0.10)$, atoms distribute randomly. As for higher concentrations, atoms gradually form an orderly arrangement, leading to the emergence of a superlattice.
Of the intercalation series investigated, stoichiometric ratios 1/4 and 1/3 are identified with distinct structural characteristics.
Corresponding atomic arrangements are depicted in Fig.~\ref{fig:str}.
At $x = 1/4$, interlayer Fe atoms establish a $2 a_0 \times2 a_0$ superlattice within the ab plane.
At $x = 1/3$, the Fe atoms form a $\sqrt{3}a_0\times \sqrt{3}a_0$ superlattice and facilitate a space group transition from $\mathrm{P6_3 /mmc~(No.~194)}$ to $\mathrm{P6_3 22~(No.~182)}$ ~\cite{JPCC2022Evolution}.
Two types of superlattices can be observed in the LEED pattern, as shown in Figs.\ref{fig:str}(c) and \ref{fig:str}(f).
Samples with intermediate compositions ranging from $x = 0.15$ to $x = 1/4$ exhibit defective superlattice structures~\cite{JPCC2023Bridging}.
The layered crystal structure characteristics and previous magnetic property measurements~\cite{CPB2025Fe0.25NbSe2} consistently confirm the c axis as the magnetic easy axis.

\begin{figure}[tbp]
\centering
\includegraphics[width = 0.5\textwidth]{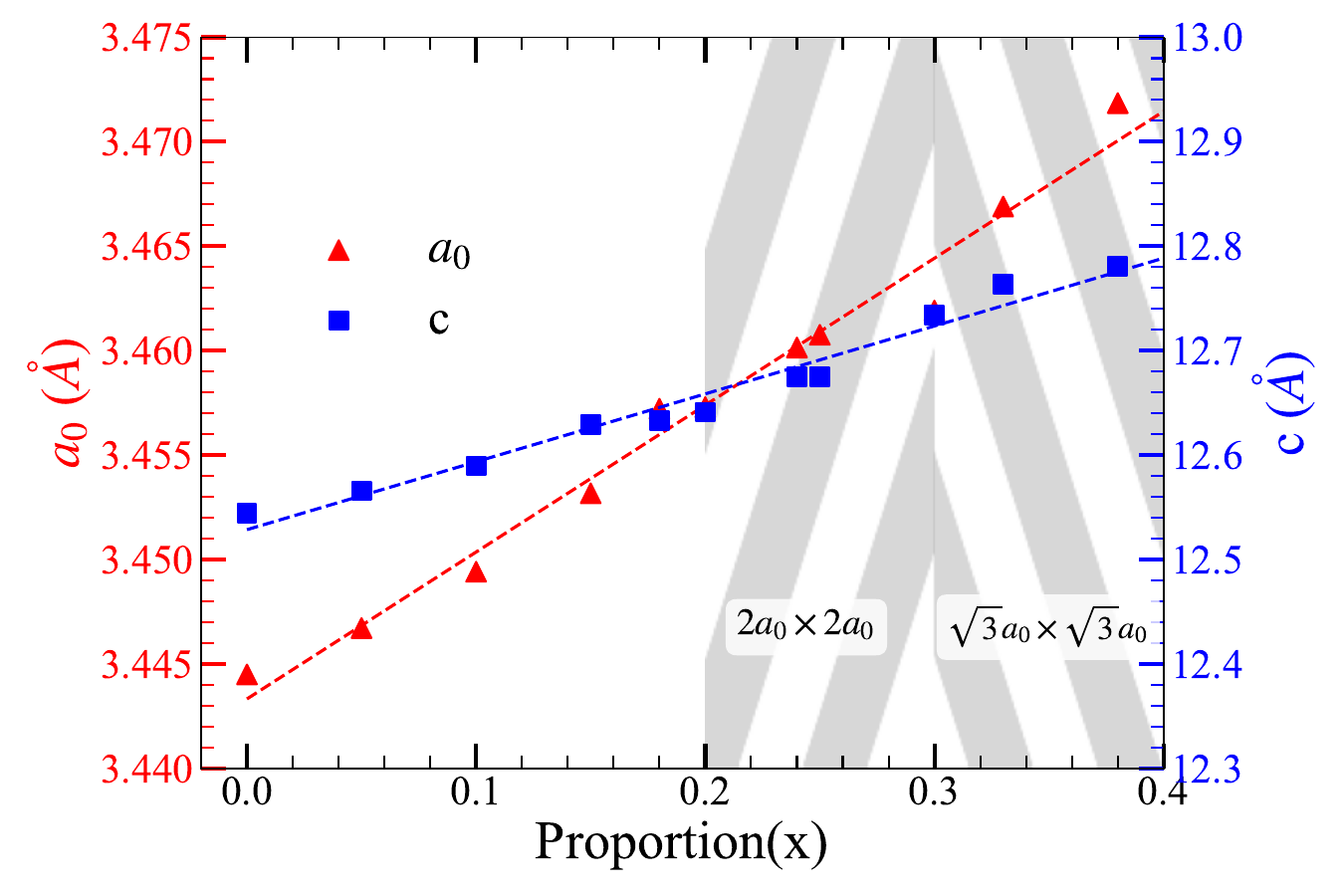}
\caption{
Evolution of the lattice parameters of \ce{Fe_{x}NbSe2} with Fe content $x$ ($0 \leq x \leq 0.38$), obtained from Rietveld refinement of powder XRD data. The shaded regions indicate the compositional ranges corresponding to the $2 a_0 \times2 a_0$ and $\sqrt{3}a_0\times \sqrt{3}a_0$ superlattice phases.
}
\label{fig:ac}
\end{figure}

The phase purity and structural parameters of the samples were confirmed by powder XRD, as shown in Fig.~\ref{fig:pxrd}. Peak positions in the patterns are found to shift toward a lower angle range.
According to the structure determined by the single crystal XRD, the formation of supercells upon Fe intercalation gives rise to new peaks, as illustrated in Fig.~\ref{fig:pxrd}(c), and typically induces changes in the lattice parameter.
Rietveld refinement results are presented in Table~\ref{tab:acr}, which demonstrated a monotonic increase in $\mathrm{a_0}$ and c-lattice parameter with increasing iron content, as depicted in Fig.~\ref{fig:ac}, implying the intercalation of Fe atoms between the $\mathrm{NbSe_2}$ layers.
The $\mathrm{R_{wp}}$ factors obtained after the final iteration of refinement are under 15\%, indicating the fitting results are reliable.

EDS, powder XRD, and LEED analyses confirmed the successful synthesis of $\mathrm{Fe}_{x}\mathrm{ NbSe_2}$ samples across varying stoichiometries, demonstrating effective Fe intercalation into the van der Waals gaps and the formation of ordered superstructures.

\begin{table}[tbp]
  \centering
  \caption{Lattice parameters $a_0$, $c$, weighted profile R-factor $\mathrm{R_{wp}}$ for the refinements and superlattice types of samples. $a_0$ in different samples refers to the parameters compared with the $\mathrm{NbSe_2}$ unit cell. The actual a-lattice parameter of the samples needs to consider the superlattice.}
  \label{tab:acr}
  \begin{tabular}{lcccc}
    \toprule
    $x$ (Fe) & $a_0$ (\AA) & $c$ (\AA) & Rwp (\%) & superlattice \\
\hline
    0.00 & 3.44450 & 12.54440 &  &  \\
    0.05 & 3.44672 & 12.56593 & 9.173 &  \\
    0.10 & 3.44942 & 12.58971 & 10.426 &  \\
    0.15 & 3.45318 & 12.62935 & 9.382 &  \\
    0.18 & 3.45722 & 12.63309 & 8.711 &  \\
    0.20 & 3.45730 & 12.64110 & 10.010 & $2 a_0 \times2 a_0$\\
    0.24 & 3.46015 & 12.67497 & 8.479 & $2 a_0 \times2 a_0$\\
    0.25 & 3.46076 & 12.67510 & 9.766 & $2 a_0 \times2 a_0$\\
    0.30 & 3.46192 & 12.73425 & 8.396 & $2a_0 \times 2a_0$, $\sqrt{3}a_0 \times \sqrt{3}a_0$ \\
    0.33 & 3.46690 & 12.76339 & 8.799 & $\sqrt{3}a_0\times \sqrt{3}a_0$\\
    0.38 & 3.47185 & 12.78053 & 9.553 & $\sqrt{3}a_0\times \sqrt{3}a_0$\\
 \hline
  \end{tabular}
\end{table}

\subsection{Intercalation effect on magnetic properties and electronic transport}\label{sec:C2}

\begin{figure*}[htbp]
\includegraphics[width=0.8\textwidth]{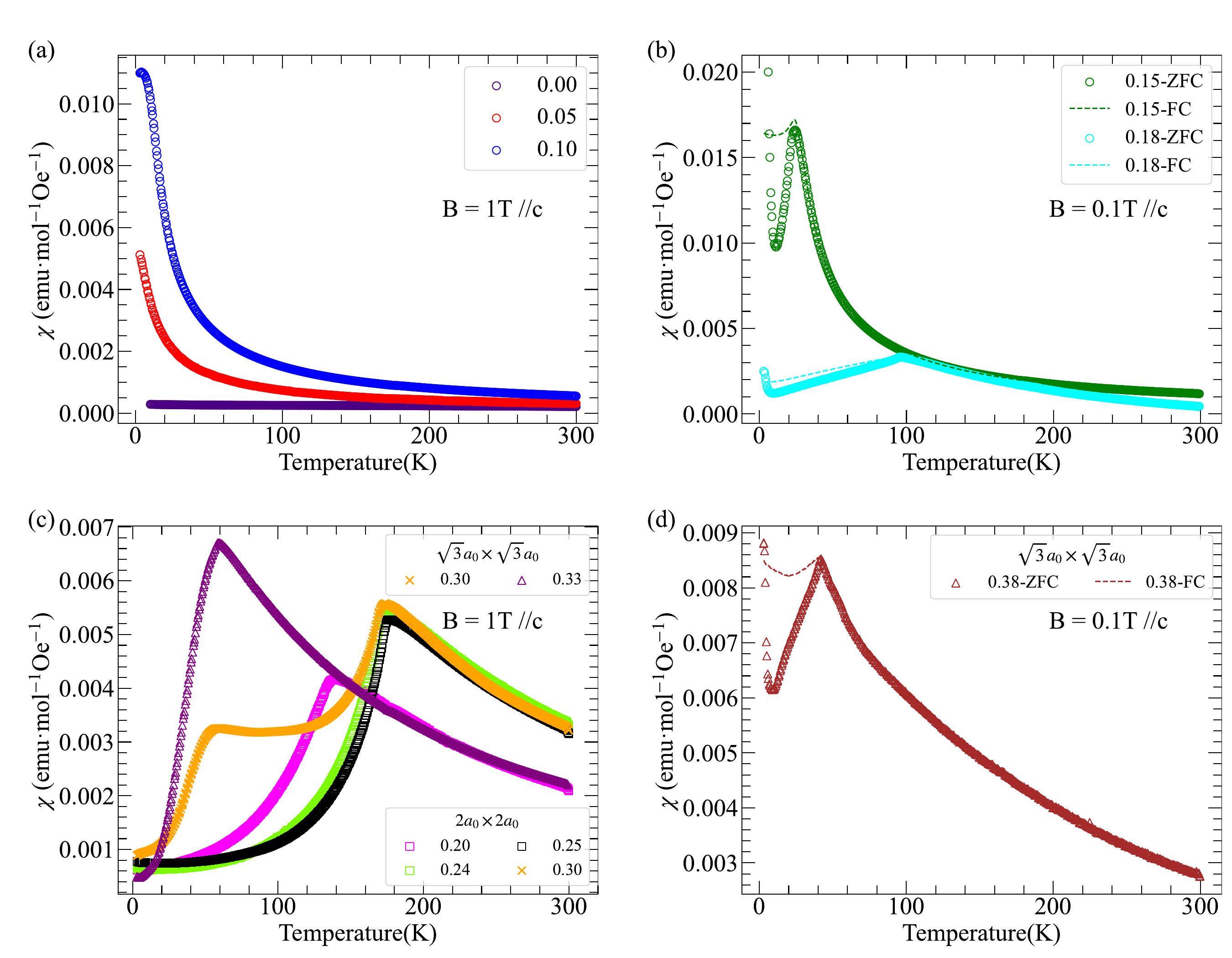}
\caption{\justifying
Temperature-dependent magnetic susceptibility $\chi(T)$ of \ce{Fe_{x}NbSe2} under a magnetic field of 1~T or 0.1~T. (a) Paramagnetic behavior at low Fe content ($x \leq 0.10$); (b) spin-glass freezing for $x = 0.15$–0.18; (c) antiferromagnetic ordering at intermediate concentrations ($x = 0.20$–0.33); and (d) reentrant spin-glass state at high doping ($x = 0.38$). ZFC and FC data are shown for selected Fe concentrations $x$.
}
\label{fig:MT}
\end{figure*}

Temperature-dependent magnetic susceptibility $\chi(T)$ of \ce{Fe_{x}NbSe2} is shown in Fig.~\ref{fig:MT}.
For samples with $ x \le 0.10$ and $0.20 \le x \le 0.33$ , only zero-field-cooled (ZFC) magnetization data are presented in Figs.~\ref{fig:MT}(a) and ~\ref{fig:MT}(c), as no bifurcation exists between the ZFC and field-cooled (FC) traces. 
Magnetic characterization reveals that Fe intercalation first suppresses the intrinsic superconductivity of \ce{NbSe2} and subsequently drives a nonmonotonic magnetic evolution: with increasing Fe content, the system transitions from a paramagnetic state to spin-glass behavior, then to long-range antiferromagnetic order, and finally back to a spin-glass state within the studied region.

For low Fe concentrations ($x = 0.05$ and $0.10$), superconductivity is fully suppressed, and the magnetic susceptibility exhibits Curie--Weiss-like paramagnetism, consistent with dilute local moments introduced by Fe intercalants. These impurity-like Fe atoms donate electrons to the host while generating localized magnetic moments, but without sufficient density to induce collective ordering.

At intermediate compositions ($x = 0.15$ and $0.18$), a clear bifurcation between ZFC and FC susceptibility curves emerges [Fig.~\ref{fig:MT}(b)], signaling the onset of spin-glass freezing---a behavior analogous to that observed in \ce{Cr_{x}NbSe2}~\cite{JAC2020CrxNbSe2} and \ce{Fe_{x}NbS2}~\cite{NJP2025FexNbS2}. 
The slight upturn in ZFC at the lowest temperatures is attributed to a Curie tail from residual paramagnetic moments.
Upon further Fe doping ($0.20 \leq x \leq 0.33$), the system develops a sharp drop in $\chi(T)$ at the N\'eel temperature $T_{\mathrm{N}}$ [Fig.~\ref{fig:MT}(c)], indicative of long-range antiferromagnetic order stabilized by commensurate Fe arrangements.

However, for $x = 0.38$, the magnetic response reverts to spin-glass behavior with Curie tail, as evidenced by ZFC--FC splitting and the absence of a well-defined transition. This resurgence suggests that excess Fe atoms occupy octahedral sites in the van der Waals gaps randomly, disrupting the periodic potential required for magnetic long-range order.

Notably, \ce{Fe_{0.30}NbSe2} exhibits two distinct antiferromagnetic transitions in its magnetic susceptibility, shown in Fig.~\ref{fig:MT}(c). The higher transition temperature coincides with the N\'eel temperature $T_{\mathrm{N}}$ observed for the $x = 0.24$ composition, which corresponds to the $2a_0 \times 2a_0$ Fe superstructure, while the lower transition matches that of the $x = 0.33$ sample associated with the $\sqrt{3}a_0 \times \sqrt{3}a_0$ ordering. The dual transitions in $x = 0.30$ suggest possible phase separation or local $\sqrt{3}a_0 \times \sqrt{3}a_0$ correlations below the XRD detection limit. However, only the $2a_0 \times 2a_0$ superlattice peaks appear in the powder XRD pattern (see Fig.~\ref{fig:pxrd}), and no signatures of the $\sqrt{3}a_0 \times \sqrt{3}a_0$ superstructure are detected, likely because it occupies a small volume fraction or produces weak scattering contrast.
Magnetic susceptibility can detect the magnetic transition associated with this minor phase due to its high sensitivity to long-range magnetic order.

\begin{table*}[t]
  \centering
  \caption{Curie-Weiss fitting parameters and superlattice information of samples. Magnetic transition temperature $T_{\mathrm{tr}}$ represents the spin-glass freezing temperature $T_{\mathrm{f}}$ or N\'eel temperature $T_{\mathrm{N}}$.}
  \label{tab:cwfit}
  \begin{tabular}{lcccccc}
    \toprule
    $x$ (Fe) & superlattice & $\mathrm{\theta_{CW}~(K)}$ & $\mathrm{C~(emu~K~mol^{-1}~Oe^{-1})}$ & $\mathrm{\mu_{eff}~(\mu_{B})}$ & $\mathrm{\mu_{eff-Fe}~(\mu_{B})}$ & $T_{\mathrm{tr}}~\mathrm{(K)}$ \\
\hline
    0.05 &  & -6.3824 & 0.0735 & 0.7670 & 3.4302 &  \\
0.10 &  & -15.4070 & 0.1746 & 1.1819 & 3.7376 & 4 \\
0.15 &  & -17.5611 & 0.3758 & 1.7339 & 4.4769 & 23 \\
0.18 &  & -22.0024 & 0.4912 & 1.9824 & 4.6725 & 97 \\
0.20 & $2 a_0 \times2 a_0$ & -34.0225 & 0.5665 & 2.1288 & 4.7601 & 135 \\
0.24 & $2 a_0 \times2 a_0$ & -79.7205 & 0.6877 & 2.3456 & 4.7879 & 172 \\
0.25 & $2 a_0 \times2 a_0$ & -83.9638 & 0.7071 & 2.3784 & 4.7569 & 175 \\
0.30 & $2a_0 \times 2a_0$, $\sqrt{3}a_0 \times \sqrt{3}a_0$  &  -72.8719 &  0.8542 &  2.6140 &  4.7726 & 172, 60 \\

0.33 & $\sqrt{3}a_0 \times \sqrt{3}a_0$ & -55.5196 & 0.9388 & 2.7405 & 4.7706 & 60 \\
0.38 & $\sqrt{3}a_0 \times \sqrt{3}a_0$ & -41.0535 & 0.9678 & 2.7826 & 4.5139 & 40 \\
\hline
  \end{tabular}
\end{table*}

\begin{figure}
\centering
\includegraphics[width = 0.48\textwidth]{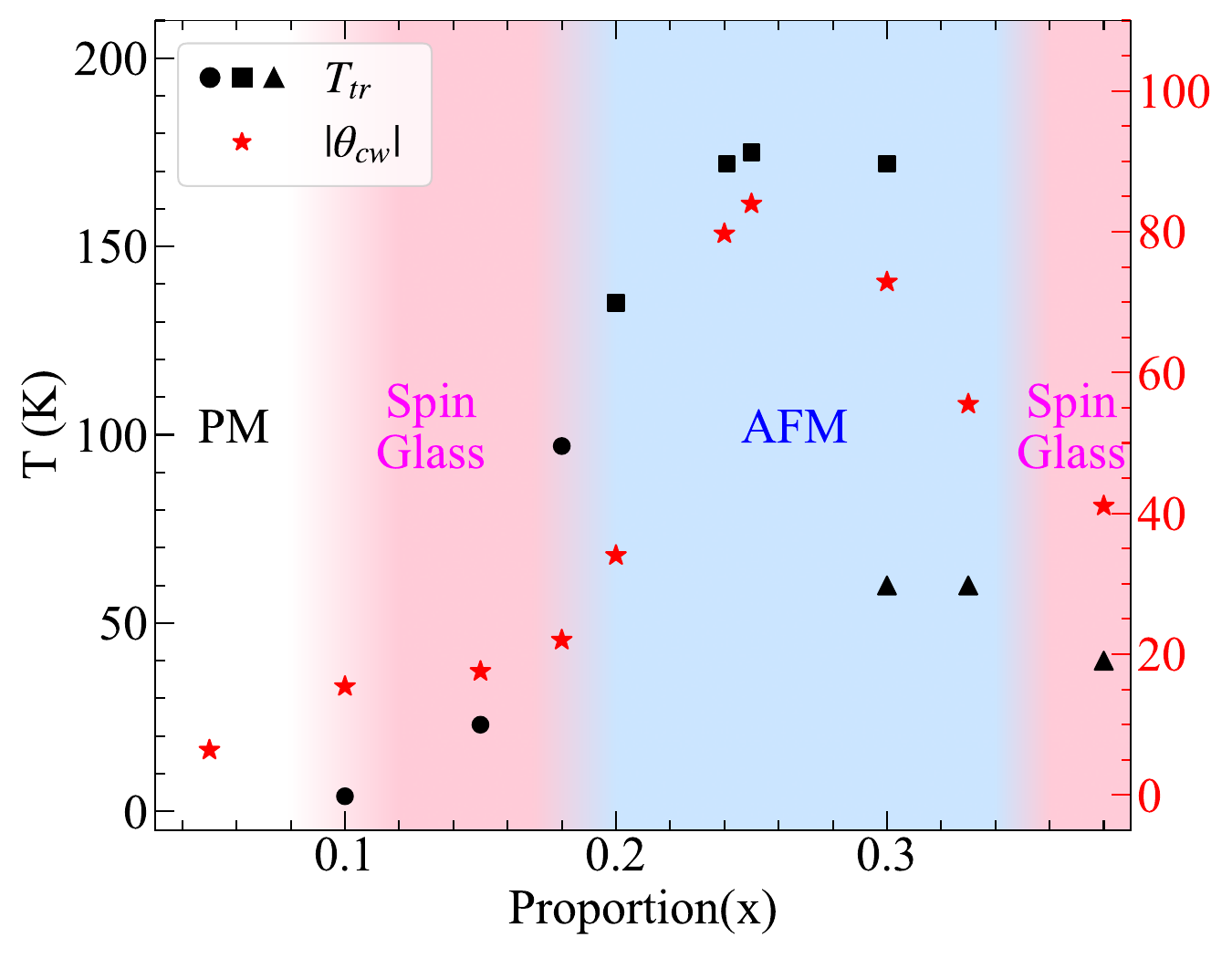}
\caption{
Magnetic phase diagram of \ce{Fe_{x}NbSe2} as a function of Fe concentration $x$. The magnetic transition temperature $T_{\mathrm{tr}}$ is represented by different symbols corresponding to distinct superlattice structures: no superlattice ($\bullet$), the $2a_0 \times 2a_0$ superlattice ($\blacksquare$), and the $\sqrt{3}a_0 \times \sqrt{3}a_0$ superlattice ($\blacktriangle$). The Curie–Weiss temperature $\theta_{\mathrm{CW}}$ from Curie–Weiss fits is shown as \textcolor{red}{$\bigstar$}. The system evolves from paramagnetic (PM) to spin-glass (SG), antiferromagnetic (AFM), and back to SG with increasing $x$.
}
\label{fig:tran}
\end{figure}

Curie--Weiss fitting parameters are summarized in Table~\ref{tab:cwfit}.
The inverse susceptibility 1/$\chi(T)$ as a function of temperature T and corresponding fitting curves are presented in the Supplemental Material (Fig. S1)~\cite{supplement}.
The effective magnetic moment ($\mathrm{\mu_{eff}}$) of Fe decreases significantly below $x = 0.15$. This suppression is likely due to increased electron delocalization and enhanced magnetic disorder in the dilute doping limit.
$\mathrm{\mu_{eff}}$ of all samples is close to $4.9\,\mu_{\mathrm{B}}$, consistent with Fe$^{2+}$. 
Both the magnetic transition temperature $T_{\mathrm{tr}}$ and the Curie--Weiss temperature $\theta_{\mathrm{CW}}$ exhibit a nonmonotonic dependence on Fe concentration $x$, as shown in Fig.~\ref{fig:tran}: they rise sharply at low $x$, peak near $x = 1/4$ at approximately 175~K, and then decrease with further Fe intercalation. This nonmonotonic trend is commonly observed in transition-metal-intercalated TMDs—including \ce{Fe_{x}NbS2}~\cite{PRB2021Fe0.33NbS2, IC2023Fe0.25NbS2}, other \ce{M_{x}NbS2} systems~\cite{PM1977MxNbS2}, and \ce{Mn_{x}TiSe2}~\cite{JAC2004MnxTiSe2}—where magnetic ordering is typically strongest near commensurate filling fractions such as $x = 1/4$ or $1/3$. 
The smooth evolution of $\theta_{\mathrm{CW}}$ reflects the gradual change in average exchange interactions with doping, while the sharp variation in $T_{\mathrm{N}}$ highlights its sensitivity to the specific Fe superlattice.

\begin{figure*}[htbp]
\includegraphics[width=0.8\textwidth]{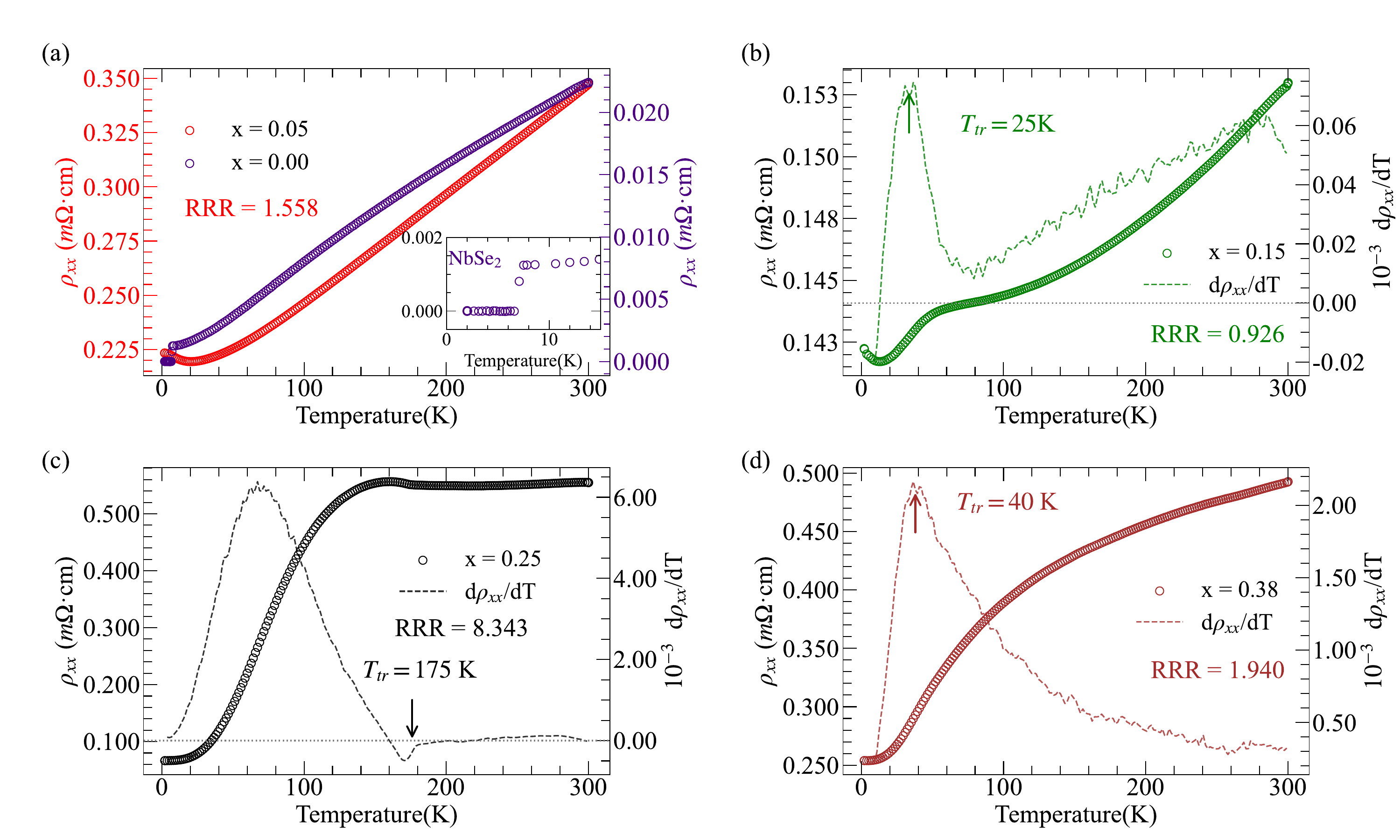}
\caption{
Electrical transport properties of samples. (a) Resistivity as a function of temperature for $\mathrm{Fe_{0.05}NbSe_2}$ and $\mathrm{NbSe_2}$. Inset shows the superconductivity of $\mathrm{NbSe_2}$. Temperature dependence of the resistivity and corresponding derivative curves of (b) $\mathrm{Fe_{0.15}NbSe_2}$, (c) $\mathrm{Fe_{0.25}NbSe_2}$, and (d) $\mathrm{Fe_{0.38}NbSe_2}$.
}
\label{fig:RT}
\end{figure*}

The evolution of magnetic ordering can be explained by the Ruderman-Kittel-Kasuya-Yosida (RKKY) interaction~\cite{PRB1997RKKY, JACS2022FeCrNbTa}. Within the dilute concentration regime, Fe atoms serve as electron donors. The increased charge carrier density~\cite{JSNM2019DFT} significantly strengthens both the magnitude and range of RKKY interactions, resulting in the rapid increase of $T_{\mathrm{tr}}$. When the concentration of Fe approaches $x = 1/4$, the intercalated Fe atoms form a highly ordered $2 a_0 \times2 a_0 $ superlattice, which determines the stable long-range magnetic order and drives $T_{\mathrm{N}}$ to its maximum value. 
For $1/4 < x \leq 1/3$, Fe begins to occupy alternative intercalation sites, eventually establishing a $\sqrt{3}a_0 \times \sqrt{3}a_0$ superstructure. This structural transformation reduces Fe-Fe atomic distance while excess Fe donates additional electrons, which collectively modify the oscillatory character of the RKKY interaction, 
shifting the RKKY interaction to a regime of weakened antiferromagnetic coupling that suppresses long-range order and leads to a sharp decline in $T_{\mathrm{N}}$.
At $x$ = 0.38, excess Fe atoms introduce disorder into the previously ordered arrangement, disrupting the long-range periodicity. Consequently, the magnetic system transitions toward a spin glass state characterized by frustrated interactions.

Additionally, electronic structure effects may further suppress magnetic ordering at high Fe concentrations. In the closely related \ce{Fe_{x}NbS2} system, theoretical and experimental studies suggest that beyond a critical concentration ($x > 1/3$), charge transfer can reverse direction, with electrons flowing back from the Nb $4d$ states to Fe $3d$ orbitals~\cite{arxiv2025FexNbS2}. This electron backflow depletes the conduction electrons in the host layers, thereby weakening the RKKY-mediated coupling or triggering Fermi surface reconstruction—both of which are detrimental to long-range magnetic order. A similar electronic instability may occur in \ce{Fe_{x}NbSe2}, where the Nb $4d$ – Se $p$ bands also serve as the primary conduction channel. The combined impact of structural disorder and such electronic modifications likely underlies the rapid collapse of $T_{\mathrm{N}}$ at high Fe concentrations, and is expected to leave distinct signatures in the electrical resistivity.

The temperature-dependent electrical resistivity $\rho(T)$ reflects the evolution of spin–electron scattering across the magnetic phase diagram, as the conduction electrons mediating the RKKY interaction are simultaneously responsible for charge transport. Representative data for $\ce{Fe_{x}NbSe2}$ spanning the four distinct magnetic regimes are shown in Fig.~\ref{fig:RT}, with features in $d\rho/dT$ (insets) closely tracking the magnetic transition temperatures $T_{\mathrm{tr}}$ identified in susceptibility measurements.
The residual resistivity ratio (RRR) reaches a maximum at $x = 1/4$, where Fe atoms form a well-ordered $2 a_0 \times2 a_0 $ superlattice, and a minimum at $x = 0.15$, corresponding to a regime of strong chemical disorder. The RRR values reflect the level of chemical disorder across the doping range.

In the dilute limit ($x = 0.05$), Fe intercalants act as localized magnetic moments that suppress superconductivity and induce a resistivity upturn at low temperatures, characteristic of Kondo scattering off paramagnetic impurities~\cite{PTP1964kondo}. At intermediate concentrations ($x = 0.15$), where spin-glass freezing occurs, the resistivity exhibits a broad feature near $T_{\mathrm{tr}}$, consistent with the onset of frozen but disordered moments that enhance inelastic scattering without establishing long-range coherence. For compositions with commensurate Fe superlattices ($x = 0.25$), the development of long-range antiferromagnetic order below $T_{\mathrm{N}}$ reduces spin-disorder scattering, giving rise to a clear kink or change in slope in $\rho(T)$—a signature of coherent magnetic ordering that sharpens in $d\rho/dT$. At higher Fe content ($x = 0.38$), the system reenters a spin-glass state due to structural and electronic disorder. Correspondingly, the resistivity shows a rounded anomaly near $T_{\mathrm{tr}}$, lacking the sharpness seen in the ordered AFM phase, yet still marking the onset of frozen spin correlations that perturb electron transport.

Magnetization and resistivity measurements across the \ce{Fe_{x}NbSe2} series reveal a unified picture: long-range magnetic order is stabilized by commensurate Fe superlattices at intermediate $x$, reaching a maximum $T_{\mathrm{N}}$ at $x = 0.25$ where the well-ordered $2a_0 \times 2a_0$ superstructure forms without excess Fe-induced disorder. At higher doping, structural and electronic inhomogeneity disrupts this periodicity, leading to the collapse of coherent magnetic order—highlighting intercalation concentration as a key knob for engineering correlated states.

\section{Conclusion}~\label{sec:E}
In conclusion, we have established a comprehensive correlation between structure, magnetism, and electrical transport in \ce{Fe_{x}NbSe2} across $0.05 \leq x \leq 0.38$. 
The magnetic transition temperature exhibits a pronounced nonmonotonic dependence on Fe concentration, peaking at $T_{\mathrm{N}} = 175~\mathrm{K}$ for $x = 0.25$, where a commensurate $2a_0 \times 2a_0$ Fe superlattice forms without chemical disorder.
As $x$ increases, the system evolves sequentially from a paramagnetic state through a spin-glass phase, into long-range antiferromagnetic order, and back to a spin-glass state at high doping, mirroring the expected behavior of an RKKY-coupled system where magnetic interactions are mediated by itinerant electrons and modulated by superlattice periodicity. 
Corresponding anomalies in the resistivity and its temperature derivative consistently track the magnetic transitions, confirming that spin–electron scattering is intimately tied to the nature of magnetic order. In intermediate compositions, multiple features in both magnetization and resistivity suggest competing or spatially inhomogeneous magnetic correlations, likely arising from partial occupancy of intercalation sites near commensurate fillings.

Notably, EDS analysis consistently revealed selenium deficiency across samples with $x \ge 0.20$, with Se:Nb ratios deviating below the ideal 2:1 stoichiometry—likely due to Se volatility during high-temperature crystal growth, a phenomenon previously observed in related transition metal dichalcogenides. Such selenium vacancies may locally perturb the electronic structure and modulate the RKKY-mediated coupling between intercalated Fe moments.

A systematic investigation of the interplay between intrinsic chalcogen vacancies and extrinsic magnetic dopants will be crucial for refining theoretical descriptions and harnessing the full potential of these materials in spintronic and quantum information applications.

\begin{acknowledgements}
The work was supported by the National Key R\&D of China under Grants No. 2022YFA1402702 and No. 2021YFA1401600, and the National Natural Science Foundation of China with Grants No. 12334008 and No. 12374148. S.G. acknowledges the financial support from the National Natural Science Foundation of China (Grants No. 22205091) and the Guangdong Pearl River Talent Plan (Grants No. 2023QN10C793). B.L. acknowledges support from the National Natural Science Foundation of China (Grants No. 12374063 and No. 92565305), the Ministry of Science and Technology of China (2023YFA1407400), and the Shanghai Natural Science Fund for Original Exploration Program (Grants No. 23ZR1479900). The experimental support from the Instrumental Analysis Center at Shanghai Jiao Tong University, especially in performing the EDS analyses.
\end{acknowledgements}

\section*{Data Availability}
The data that support the findings of this article are not publicly available. The data are available from the authors upon reasonable request.

\bibliographystyle{unsrt}
\bibliography{FNSbib}

\end{document}